\documentclass[
 reprint,
 amsmath,amssymb,
 aps, floatfix, prd
]{revtex4-2}

\usepackage{graphicx}
\usepackage{dcolumn}
\usepackage{bm}
\usepackage{aas_macros}
\usepackage{hyperref}
\usepackage{xcolor}
\usepackage{multirow}
\usepackage{ulem}
\usepackage{bm}

\begin{document}

\preprint{XXX}

\title{Constraining ultralight bosons in dwarf spheroidal galaxies with a radially varying anisotropy}

\author{Ivan de Martino}%
\email{ivan.demartino@usal.es}
\affiliation{%
    Departamento de F\'isica Fundamental and IUFFyM, Universidad de Salamanca,\\Plaza de la Merced, s/n, E-37008 Salamanca, Spain
}%

\date{\today}%

\begin{abstract} 
Axions, and axion-like particles, have come back into fashion in the last decades as a possible solution to the galactic-scale crisis suffered by the cold dark matter model. In the framework of the wave Dark Matter model, we have carried out a Jeans Analysis on eight dwarf spheroidal galaxies that are orbiting around the Milky Way, and we have constrained the boson mass. Differently to a previous analysis, we adopted an anisotropy parameter that varies with the distance from the centre of the galaxy to assess whether this assumption would help to resolve, or at least alleviate, the well-known tension with the value of the boson mass favoured by the cosmological analysis. Our results indicate that, differently to what happens in ultra-faint dwarf galaxies, such a tension cannot be lifted introducing a variable anisotropy parameter, leaving as a possible solution the existence of additional axion or axion-like particles with higher masses as naturally predicted in the Axiverse.
\end{abstract}

\maketitle

\section{Introduction}\label{sec:intro}

{ Although the Cold Dark Matter (CDM) model successfully describes the cosmological evolution of the Universe, it faces unrelieved challenges on the scales of galaxies. These issues are mainly related to the dark matter distribution in the innermost regions of the galactic halos. For instance, dwarf galaxies seem to prefer a cored dark matter mass density profile in contrast to the cuspy profile predicted by N-body simulation in the CDM context \cite{Navarro1996}. This is the well-known cusp-core problem which, to some extent, can be alleviated by the baryonic feedback \cite{idm2020Univ}. However, morphological features related to tidal disruption of ultra-diffuse dwarf galaxies in the Fornax Cluster have been recently shown to be incompatible with expectations from CDM  at $M^* < 10^{7.2} M_\odot$ where baryonic feedback would not play any role \cite{Asencio2022}.  Additionally, the dynamical friction of the dark matter halo should slow down the galactic bars as predicted from cosmological simulations. However, observed galaxies pointed out $12.6\sigma$ tension with the simulated ratio of bar length to the radius of
corotation with the bar pattern speed \cite{Roshan2021}. Both issues could be alleviated if the central dark matter mass density was reduced with respect to the one predicted in the CDM model. One alternative that could provide such a mass density profile is given by ultralight particles.}

Firstly introduced by \cite{Widrow1993, Sahni2000, Hu2000}, ultralight bosons with masses ranging from $10^{-24}$ eV to $10^{-17}$ eV have come back into fashion in the last decades to overcome the galactic-scale issues pointed out in the context of CDM \cite{Marsh2014, Schive2014a, Bozek2015, Hui2017, Veltmaat2018, Robles2019, idm2020Univ, Ferreira2021, Cesare2023}. The most promising candidates are axions or axion-like particles naturally produced in the String theory landscape \cite{Arvanitaki2010, Marsh2014}. In this model, which is usually referred to as fuzzy, wave, or $\psi$ Dark Matter (DM), when self-interaction is not taken into account, the only free parameter is the mass of the particle ($m_\psi)$. However, there is still no unanimous agreement on its fiducial value. Indeed, there exists a strong tension between the boson mass of the order of $10^{-22}$ eV preferred by galactic kinematics, and the boson mass 
of the order of $10^{-20}$ eV preferred by the cosmological evolution of the matter in the Universe (for a comprehensive review we refer to \cite{Marsh2015, Marsh2016, Hui2017, Niemeyer2020, Ferreira2021, Hui2021}).

More in detail, a boson mass $m_\psi\sim 10^{-22}$ eV allows us to explain in a single framework several observations on the galactic scale. In \cite{idm2020}, the excess in the central velocity dispersion of stars residing in the bulge of the Milky Way can be explained with the presence of a dark solitonic core of mass $\sim 10^9 M_\odot$ originated by the gravitational collapse of ultralight bosons having a mass $m_\psi =(0.9\pm0.06)\times 10^{-22}$ eV. In \cite{Chen2017}, the Jeans analysis carried out on eight dwarf spheroidal (dSph) galaxies orbiting the Milky Way, allowed to fit the velocity dispersion profiles with a common boson mass of $m_\psi = 1.79^{+0.35}_{-0.33}\times10^{-22}$ eV at the 95\% of the confidence interval. The large size galaxy with slowly moving stars, namely Antlia II, is also explained with a boson mass of $m_\psi = 0.81^{+0.41}_{-0.21}\times10^{-22}$ eV at the 68\% of confidence interval \cite{Broadhurst2020}. Finally, a boson mass of $m_\psi = 2.1^{+4.9}_{-1.3}\times10^{-22}$ eV can account for the relatively flat velocity dispersion profile of `Dragon Fly 44' \cite{Wasserman2019,Pozo2021}, { though alternative explanations relying on Modified Newtonian Dynamics and Modified Gravity may also account for the observed line-of-sight velocity dispersion profile with mass-to-light ratio in agreement with stellar population synthesis models \cite{Haghi2019}.}
Although probes relying on the kinematics of stars in galaxies seem to favour a boson mass of $m_\psi\sim 10^{-22}$ eV, some analyses question those results. For instance, wave DM with a boson mass of $m_\psi = 2.5^{+3.6}_{-2.0}\times10^{-21}$ eV might explain the observed rotation curve of Milky Way but it would introduce a $\sim 5\sigma$ tensions with the previous results on dwarf galaxies \cite{Maleki2020}. Additionally, explaining the density profile of ultra-faint dwarf galaxies would require a boson mass of $\gtrsim 10^{-21}$ eV \cite{Safarzadeh2020} which, however, should be ruled out by the density profile of the classical dwarves. Even on the scale of galaxies, there is no unanimous agreement between different datasets and methodologies. { On the other hand, the comoving cumulative stellar mass density of massive galaxies at $7 < z < 11$, which have been measured in the JWST early data release, would require, to be explained in the CDM model, high star formation efficiency that, on turn, would increase the number of ionizing photons generating tension with cosmic microwave background and measurements of cosmic reionization history. This would be partially avoided in the wave DM scenario with a boson mass $m_\psi \sim 10^{-22}$ eV \cite{Gong2023}.}

On a much smaller scale, relying on the observations of the supermassive black hole in M87 made by the Event Horizon Telescope Collaboration \cite{EHT2019}, analysis modelling the superradiance instability gives two allowed regions for the boson mass:  $m_\psi\lesssim2.9\times 10^{-22}$ eV and $m_\psi\gtrsim4.6\times 10^{-22}$ eV \cite{Davoudiasl2019}, { excluding a narrow range of masses while still leaving untouched much of the parameter space.} More recently,  astrometric and spectroscopic observations of S2 star orbiting the supermassive black hole at the centre of the Milky Way have been used to set an upper limit on the boson mass equal to $m_\psi = 3.2\times 10^{-19}$ eV at 95\% confidence interval \cite{DellaMonica2023a,DellaMonica2023b}. Those results can easily agree with the constraints on both the galactic and cosmological scales. 

Cosmological analyses can constrain the value of the boson mass in a wide range from $10^{-24}$ eV to $10^{-17}$ eV. Analysis using the observations of the cosmic microwave background and large-scale structure are not very sensitive to the value of the boson mass which, indeed, must be $>10^{-24}$ eV \cite{Hlozek2015, Hlozek2018}. Nevertheless, analyses that use the Lyman-$\alpha$ forest data got a much better constraint setting the lower limit to the boson mass to $m_\psi \gtrsim 2\times10^{-20}$ eV at the 95\% of the confidence interval \cite{Rogers2021}. While those results are in strong tension with the aforementioned results on the galactic scale, interestingly, another recent result obtained carrying out a Jeans analysis of ultra-faint galaxies that employs a variable anisotropy parameter seems also to favour a boson mass heavier than $\times 10^{-21}$ bringing galactic and cosmological observations into agreement. Indeed, in such a model, the kinematic of the star in galaxies would require a boson mass of $m_\psi=1.1^{+8.3}_{-0.7}\times 10^{-19}$ eV to fit the velocity dispersion profile of the ultra-faint dwarf galaxy Segue 1 \cite{Hayashi2021}.  

Recently, high-resolution N-Body simulations that have been used to predict the dynamical heating of dwarf galaxies in a wave DM halo with a boson mass $m_\psi= 8\times 10^{-22}$ eV, also have shown that the sizes of the dwarf galaxies and their central velocity dispersion increase over time, and their velocity distribution becomes radially anisotropic in the outskirts favouring, in fact, a variable anisotropy parameter also in dwarf galaxies \cite{Dutta2023}.
Then, our main objective is to investigate whether adopting a variable anisotropy parameter in the Jeans analysis of dSph galaxies can help to at least alleviate, if not resolve, the tension between this data set and cosmological ones. In Section \ref{sec:model}, we give the details of the theoretical model used to fit the velocity dispersion profiles of the dSph galaxies. Then, in Sect. \ref{sec:data}, we move to summarize the statistical methodology and the dataset used to constrain the model. Finally, in Sect. \ref{sec:results} and \ref{sec:concl}, we expose our results and the conclusions, respectively.

\section{Modelling the velocity dispersion profile}\label{sec:model}

\subsection{The Jeans equation} 

The kinematic of stars in dwarf galaxies is driven by the gravitational potential well of the DM halo, being the latter the primary mass component, { while their velocity dispersion can reliably be predicted from the visible mass \cite{McGaugh2021,Banik2022}}. dSph galaxies can be modelled as spherically symmetric self-gravitating systems  in dynamical equilibrium, supported by the velocity dispersion, through the spherically symmetric  Jeans equation \cite{Lokas2003,Mamon2005,Binney2008,Mamon2010} 
 \begin{equation}\label{eq:Jeans}
 	\frac{d[\nu_*(r)\sigma_r^2(r)]}{dr} + 2{  \beta(r)}\frac{\nu_*(r)\sigma_r^2(r)}{r} = -\nu_*(r)\frac{GM(r)}{r^2}\,.
 \end{equation}

In the above equation, $\nu_*(r)$ is the number density profile of the tracing stellar population, and $\sigma_r(r)$ is the radial component of the velocity dispersion. The velocity anisotropy parameter $\beta(r)$ can be defined as 
\begin{equation}\label{eq:beta_def}
 \beta(r) \equiv 1-\frac{\sigma_t(r)}{2\sigma_r(r)} \,,
\end{equation}
where $\sigma_t(r)$ is the tangential component of the velocity dispersion. Usually, the velocity anisotropy parameter is taken as constant with $\beta<0$ ($\beta>0$) indicating a tangentially (radially) biased velocity distribution, and $\beta=0$ an isotropic velocity distribution. Nevertheless, it depends on the radius by definition and, hereafter, we will retain the radial dependence to test its impact on the constraints of the boson mass. Finally, $M(r)$ is the enclosed mass of the DM halo. The general solution to the  equation \eqref{eq:Jeans} is
\begin{equation}
\nu(r)\sigma_r=F^{-1}(r)\int_r^{\infty}F(s)\nu(s)\frac{GM(s)}{s^2}ds\,,
\end{equation}
where
\begin{equation}
F(r)=\exp\biggl(\int_0^r 2\frac{\beta(t)}{t}dt\biggr).
\end{equation}
Nevertheless, such a solution must be projected along the line of sight to fit the model to the data. The projection is achieved as follows
\begin{equation}\label{eq:sigmalos}
\sigma_{los}^2(R)=\frac{2}{\Sigma(R)}\int_{R}^{\infty}\biggl (1-\beta(r)\frac{R^2}{r^2}\biggr ) \
              \frac{\nu(r)\sigma_rr}{\sqrt{r^2-R^2}}dr,
\end{equation}
where $R$ is the projected radius, $\Sigma(R)$ is the stellar surface density, and $\sigma_{los}^2(R)$ is line-of-sight velocity dispersion.

\subsection{The stellar number density profile}

To trace the stellar surface density of dSph galaxies, we follow \cite{Walker2009d} and adopt a Plummer profile:
\begin{equation}\label{eq:2DPlummer}
    I(R)=L(\pi r_h^2)^{-1}[1+R^2/r_h^2]^{-2}\,.
\end{equation}
Here $L$ and $r_h$ represent the total luminosity in a given observational band and the radius enclosing half of the total luminosity, respectively. 
The assumption of a constant mass-to-light ratio allows us to derive the three-dimensional density profile from the equation \eqref{eq:2DPlummer} through the Abel transform as
\begin{equation}
\nu(r)=3L_V(4\pi r_h^3)^{-1}[1+r^2/r_h^2]^{-5/2}.    
\end{equation}

Finally, we adopt the luminosity in the $V$-band ($L_V$), and the half-light radius ($r_h$)  from the observations in \cite{Walker2009d} that are also reported, for the sake of completeness, in Table \ref{tab:0}.
	\begin{table}[!ht]
		\begin{center}
			\resizebox{7cm}{!}{
				\setlength{\tabcolsep}{4pt}
				\begin{tabular}{lcc}
					\hline
					\hline
					Galaxy &$\log_{10} \biggl(\dfrac{L_{\rm V}}{L_\odot}\biggr)$  & $r_h$   \\
					&                 & (pc)    \\
					\hline
					\textbf{Carina} & 5.57$\pm$0.20 & 273$\pm$45 \\[0.1cm]
					\textbf{Draco} & 5.45$\pm$0.08 & 244$\pm$9 \\[0.1cm]
					\textbf{Fornax} & 7.31$\pm$0.12 & 792$\pm$58 \\[0.1cm]
					\textbf{Leo I} &  6.74$\pm$0.12 & 298$\pm$29\\[0.1cm]
					\textbf{Leo II} &  5.87$\pm$0.12 & 219$\pm$52 \\[0.1cm]
					\textbf{Sculptor} & 6.36$\pm$0.20 & 311$\pm$46 \\[0.1cm]
					\textbf{Sextans} & 5.64$\pm$0.20 & 748$\pm$66\\[0.1cm]
					\textbf{Ursa Minor} & 5.45$\pm$0.20 & 398$\pm$44 \\[0.1cm]
					\hline
				\end{tabular}
			}
		\end{center}
	\caption{Observational properties of the eight dSph galaxies analysed in this work. The first column lists the names of the galaxies. The second and third columns report the total $V$-band luminosity and the half-light radius, respectively.}\label{tab:0}
	\end{table}

\subsection{The wave Dark Matter halo}

Ultralight bosons without self-interaction have been the subject of a deep investigation as an alternative candidate to cold dark matter. Pioneering N-body simulations
in this context brought to light the existence of a solitonic stationary core in each virialized halo, which is the ground-state solution of the coupled Schr\"oedinger-Poisson equations.  The solitonic core is well { fitted} by the following { formula} \cite{Schive2014a, Schive2014b}
\begin{align}
\rho_{sol}(r) & \sim \biggl[1.9~a^{-1} \biggl(\frac{m_\psi}{10^{-23}~{\rm eV}}\biggr)^{-2}\biggl(\frac{r_c}{\rm kpc}\biggr)^{-4}\biggr]\times \nonumber\\
& \biggl[1+9.1\times10^{-2}\biggl(\frac{r}{r_c}\biggr)^2\biggr]^{-8} ~M_\odot {\rm pc}^{-3}\,. \label{eq:sol_density}
\end{align}
Here $a$ is the cosmological scale factor, and $r_c$ is the radius of the solitonic core whose size scales as a power of the halo mass \citep{Schive2014b}, { and is inversely proportional to the mass of the boson $m_\psi$, hence, lower mass implies a larger solitonic core radius}. Nevertheless, following \cite{Chen2017}, we will treat $r_c$ as an independent free parameter.

Such a solitonic core is surrounded by a wave-like interference pattern modulated on the de Broglie scale \cite{Schive2014a}. Such a modulation reflects on the amplitude of the Compton frequency oscillation of the bosonic scalar field whose direct detection could be achieved with the future-generation radio telescopes ({\textit{e.g.}, Square Kilometer Array (SKA)}) \citep{idm2017,idm2018, Luu2023}. Surprisingly, the azimuthal average of such an extended halo region follows the Navarro-Frank-White (NFW) density profile \citep{Navarro1996, Schive2014a} and, hence, the total DM density profile of the whole halo may be written as follows:
\begin{equation}\label{eq:dm_density}
\rho_{DM}(r) =
\begin{cases} 
\rho_{sol}(x)  & \text{if \quad}  r\leq r_t\,, \\\\
\dfrac{\rho_0}{\dfrac{r}{r_s}\bigl(1+\dfrac{r}{r_s}\bigr)^2} & \text{if \quad}   r> r_t\,.
\end{cases}
\end{equation} 
Here, the transition radius $r_t$ from one regime to the other is set to $3\,r_c$ on the basis of the N-body simulations \cite{Schive2014a}, $\rho_0$ is set to match the inner (solitonic) and outer (NFW-like) profiles at $r_t$, and, finally $r_s$ is the scale radius.

\subsection{The radial-dependent anisotropy parameter}\label{subsec:betas}

Recent numerical simulations \cite{Dutta2023} of the dynamical heating of dwarf galaxies embedded in a wave DM halo have shown that the initial isotropic velocity distribution becomes radially anisotropic over time following an Osipkov-Merritt model \cite{Osipkov1979,Merritt1985}. Additionally, a Jeans analysis that incorporates an anisotropy parameter variable with the distance from the centre of the galaxy, is capable of both correctly predicting the velocity dispersion profiles of ultra-faint dwarves, and relieving the tension with cosmological analysis \cite{Hayashi2021}. However, a similar analysis with dSph galaxies is missing, and it is interesting to investigate the impact of a variable velocity anisotropy distribution on the constrained boson mass. 

To our aims, we consider three possible scenarios: velocity anisotropy-density slope (hereafter VADS) relation model \cite{Hansen2006, Hansen2006b, Zait2008};  the Osipkov-Merritt (hereafter OM) model \cite{Osipkov1979,Merritt1985}, and the generalization of the Osipkov-Merritt  (hereafter gOM) model \cite{Baes2007}:
\begin{align}
\text{\bf VADS Model}\quad &\beta(r) = 1 - 1.15\biggl[ 1+ \frac{1}{6}\frac{r}{\rho(r)}\frac{d\rho(r)}{dr}\biggr]\,,\\
\text{\bf OM Model}\quad &\beta(r) = \frac{r^2}{r^2+r_a^2}\,,\\
\text{\bf gOM Model}\quad &\beta(r) = \beta_0 + (\beta_\infty - \beta_0)\frac{r^2}{r^2+r_\beta^2}\,.
\end{align}
where $r_a$ and $r_\beta$ are the so-called anisotropy radius. 

\begin{figure*}[!ht]
    \centering
    \includegraphics[width= 2\columnwidth]{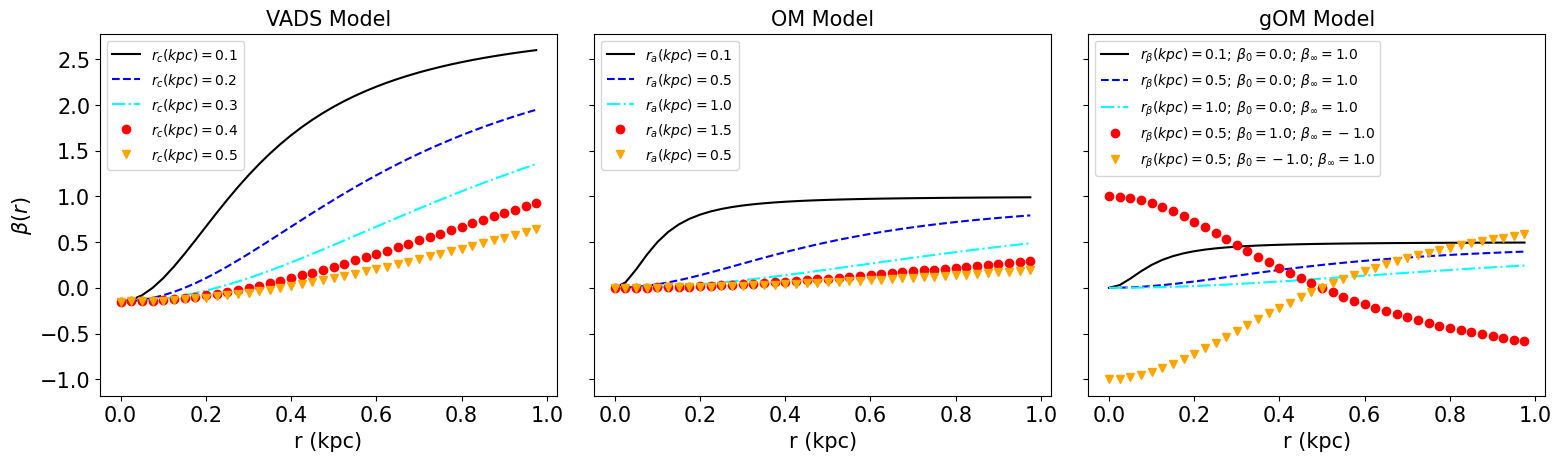}
    \caption{The figure depicts the VADS, OM, and gOM models in the left, middle, and right panels, respectively. The values of the parameters chosen to draw the models are listed in the legend in each panel. }
    \label{fig:betas_profiles}
\end{figure*}

In Figure \ref{fig:betas_profiles}, we depicted the trend of the VADS, OM, and gOM models in the left, middle, and right panels, respectively.
In the left panel, the VADS model reflects the results of N-body simulations that identify a universal relation between the slope of the mass density profile of the halo, and the velocity anisotropy distribution. It depends explicitly only on the core radius of the wave DM halo, and it always tends to a more radially anisotropic velocity distribution. 
In the middle panel, it is shown the OM model whose velocity distribution is built to be almost radial ({\textit{e.g.}}, $\sigma_r\gg \sigma_t)$ for  $r\gg r_{a}$ and nearly isotropic (\textit{e.g.}, $\sigma _{r}\simeq \sigma _{t}$) for $r\ll r_{a}$, which is rather expected since self-gravitating stellar systems have isotropic and radially-anisotropic velocity distributions in the innermost and outermost regions of the galaxy, respectively. Finally, in the right panel, it is shown the gOM model that aims to describe velocity distributions with a variable anisotropy parameter ranging from a $\beta_0$ in the innermost regions to $\beta_\infty$ in the outermost regions, without a priori limitations on their values \cite{Albada1982, Dutta2023}.

\section{Data and Methodology}\label{sec:data}

We solve the Jeans equation in \eqref{eq:Jeans} adopting the wave DM halo to predict the theoretical velocity dispersion profile and, then, we computed the projected velocity dispersion profile along the line of sight ($\sigma_{\mathrm{los,\, th}}(r)$) as shown in equation \eqref{eq:sigmalos}. Finally, we 
use the measured line-of-sight velocity dispersion profiles  ($\sigma_{\mathrm{los,\, obs}}(r)$) of eight dSph galaxies, and  their observational uncertainties ($\Delta\sigma_{\mathrm{los,\, obs}}(r_i)$),
to constrain the wave DM halo parameters and the parameters associated with the specific model of the velocity distribution.

To carry out our analysis, we employ an MCMC algorithm to explore the parameter space $\bm{\theta}$. Specifically, we use the \texttt{emcee} Python package \cite{emcee}. Then, on all the free parameters, we set uniform prior distributions as reported in Table \ref{tab:tab1}.
\begin{table}[!ht]
    \centering
    \resizebox{8.5cm}{!}{
				\setlength{\tabcolsep}{4pt}
    \begin{tabular}{cc||cc}
    \hline
    \hline
        \textbf{Parameter} & \textbf{Prior} & \textbf{Parameter} & \textbf{Prior} \\
        \hline
        & & & \\[-0.1cm]
         $\log_{10} \biggl(\dfrac{m_\psi}{\text{eV}}\biggr)$ &  $\mathcal{U}(-24, 17)$ & $\beta_0$ & $\mathcal{U}(-10, 1)$\\[0.4cm]
         $\dfrac{r_c}{\text{kpc}}$ & $  \mathcal{U}(0, 3)$ & $\beta_\infty$ & $\mathcal{U}(-10, 1)$\\[0.4cm]
         $\dfrac{r_a}{\text{kpc}}$ & $  \mathcal{U}(0, 10)$ &&\\[0.4cm]
         \hline
         \hline
    \end{tabular}
    }
    \caption{Prior distributions of the model parameters.}
    \label{tab:tab1}
\end{table}

Then, we run a total number of chains at least two times larger than the number of dimensions of the parameter spaces, and we let them run until the convergence is reached. Two conditions must be satisfied by the chains to ensure the convergence:  the first one is that the length of each chain must be 100 times longer than the autocorrelation time and, the second one is that the autocorrelation time must change by less than 1\%  between two consecutive checks (for more details we refer to Sec. 3 of \cite{deMartino2022}). Finally, the log-likelihood distribution is computed as follows
    \begin{align}
    	-2\ln \mathcal{L}(\bm{\theta}|\textrm{ data}) \propto& \sum_i\biggl[\frac{\sigma_{\mathrm{los,\, th}}(\bm{\theta},\, R_i)-\sigma_{\mathrm{los,\, obs}}(R_i)}{\Delta\sigma_{\mathrm{los,\, obs}}(R_i)}\biggr]^2\,.
    	\label{eq:likelihood}
    \end{align}

The kinematic data $\sigma_{\mathrm{los,\, obs}}$ of the eight dSph galaxies were obtained with the Michigan/MIKE Fiber Spectrograph in the case of Carina, Fornax, Sculptor, and Sextans \cite{Walker2007, Walker2009a, Walker2009b, Walker2009c, Walker2009d}, and with the Hectochelle fiber spectrograph at the MMT in the case of Draco, Leo I, Leo II, and Ursa Minor  \cite{Mateo2008}. Additionally, for each galaxy, the values of the luminosity in the $V$-band, and the half-light radius are taken from \cite{Walker2009d}.

In the case of the dSph galaxies, the transition scale, $r_t$, from the solitonic to the NFW-like regime, is much larger than the half-light radius making that all the observed stars (used in the analysis) reside within the soliton. Therefore, for the sake of simplicity, we will model the wave DM halo only using the solitonic core. It has been shown in \cite{Chen2017} that, even in the cases where this approximation could fail, 
accounting in the analysis for the outer region of the DM halo neither improves nor changes the final constraints on the boson mass. Therefore, the parameter space of the DM halo only comprises the boson mass $m_\psi$ and the core radius $r_c$.

\section{Results and Discussions}\label{sec:results}

\begin{table*}[!ht]
    \setlength{\tabcolsep}{15pt}
    \renewcommand{\arraystretch}{1.5}
\begin{center}
\begin{tabular}{ccccccc}
   \hline
    \textbf{Galaxy} & $\beta$-Model   & $r_c$ (kpc) & $\log_{10}  m_\psi$ &  $r_a$ (kpc) & $\beta_0$ & $\beta_\infty$\\
    \hline
     \multirow{3}{*}{\textbf{Carina}} & 
       VADS Model& $0.40^{+0.05}_{-0.04}$ & $-21.45\pm0.07$ & ... & ...  & ... \\
     & OM Model & $0.43^{+0.17}_{-0.08}$ & $-21.53^{+0.14}_{-0.20}$ & $1.57^{+1.02}_{-1.08}$ & ...  & ... \\
     & gOM Model & $0.47^{+0.15}_{-0.12}$ & $-21.59^{+0.19}_{-0.18}$ & $5.45^{+3.07}_{-3.04}$ & $0.32^{+0.18}_{-0.26}$  & $-4.33^{+3.71}_{-3.79}$ \\
    \hline
    \multirow{3}{*}{\textbf{Draco}} & 
       VADS Model& $0.68^{+0.04}_{-0.03}$ & $-21.99\pm0.04$  & ... & ...  & ...  \\
     & OM Model & $0.52^{+0.11}_{-0.07}$ & $-21.8^{+0.10}_{-0.11}$ & $1.87\pm 0.74$ & ...  & ... \\
     & gOM Model & $0.58^{+0.11}_{-0.12}$ & $-21.93^{+0.16}_{-0.12}$ & $5.23^{+3.16}_{-2.81}$ & $0.54^{+0.21}_{-0.32}$  & $-4.85^{+3.89}_{-3.59}$ \\
    \hline
    \multirow{3}{*}{\textbf{Fornax}} & 
       VADS Model& $0.65^{+0.03}_{-0.01}$ & $-21.75\pm0.02$  & ... & ...  & ... \\
     & OM Model & $0.73^{+0.32}_{-0.10}$ & $-21.88^{+0.08}_{-0.20}$ & $1.53^{+0.98}_{-0.69}$ & ...  & ... \\
     & gOM Model & $0.58^{+0.07}_{-0.08}$ & $-21.77^{+0.08}_{-0.06}$ & $6.08^{+2.73}_{-3.68}$ & $0.07^{+0.11}_{-0.12}$  & $-3.04^{+2.95}_{-4.41}$ \\
    \hline
    \multirow{3}{*}{\textbf{Leo I}} & 
       VADS Model& $0.40^{+0.04}_{-0.03}$ & $-21.62^{+0.05}_{-0.06}$  & ... & ...  & ... \\
     & OM Model & $0.43^{+0.30}_{-0.08}$ & $-21.70^{+0.12}_{-0.34}$ & $1.30^{+1.16}_{-0.97}$ & ...  & ... \\
     & gOM Model & $0.42^{+0.14}_{-0.11}$ & $-21.69^{+0.18}_{-0.20}$ & $5.60^{+2.92}_{-3.49}$ & $0.36^{+0.27}_{-0.44}$  & $-4.22^{+3.84}_{-3.90}$ \\
    \hline
    \multirow{3}{*}{\textbf{Leo II}} & 
       VADS Model& $0.18^{+0.03}_{-0.02}$ & $-21.09^{+0.07}_{-0.10}$  & ... & ...  & ... \\
     & OM Model & $0.19^{+0.30}_{-0.06}$ & $-21.18^{+0.18}_{-0.62}$ & $0.90^{+1.39}_{-0.81}$ & ...  & ... \\
     & gOM Model & $0.23^{+0.20}_{-0.09}$ & $-21.31^{+0.28}_{-0.41}$ & $5.55^{+3.01}_{-3.18}$ & $0.60^{+0.29}_{-0.64}$  & $-4.21^{+3.72}_{-3.82}$ \\
    \hline
    \multirow{3}{*}{\textbf{Sculptor}} & 
       VADS Model& $0.43^{+0.03}_{-0.03}$ & $-21.64^{+0.03}_{-0.04}$  & ... & ...  & ... \\
     & OM Model & $0.48^{+0.46}_{-0.10}$ & $-21.76^{+0.14}_{-0.43}$ & $0.91^{+1.52}_{-0.61}$ & ...  & ... \\
     & gOM Model & $0.41^{+0.07}_{-0.06}$ & $-21.65\pm{0.10}$ & $5.36^{+3.13}_{-3.07}$ & $0.14^{+0.13}_{-0.16}$  & $-4.42^{+3.70}_{-3.80}$ \\
    \hline
    \multirow{3}{*}{\textbf{Sextans}} &
       VADS Model& $0.53^{+0.19}_{-0.08}$ & $-21.39^{+0.12}_{-0.24}$  & ... & ...  & ... \\
     & OM Model & $0.65^{+0.30}_{-0.17}$ & $-21.59^{+0.16}_{-0.15}$ & $1.67^{+0.82}_{-0.68}$ & ...  & ... \\
     & gOM Model & $0.25^{+0.20}_{-0.11}$ & $-21.17^{+0.19}_{-0.26}$ & $5.33^{+3.15}_{-3.42}$ & $-0.85^{+0.75}_{-2.41}$  & $-4.33^{+3.66}_{-3.85}$ \\
    \hline
    \multirow{3}{*}{\textbf{UMi}} & 
       VADS Model& $0.39^{+0.07}_{-0.05}$ & $-21.56^{+0.09}_{-0.12}$  & ... & ...  & ... \\
     & OM Model & $0.45^{+0.16}_{-0.09}$ & $-21.69^{+0.16}_{-0.20}$ & $1.61\pm 0.94$ & ...  & ... \\
     & gOM Model & $0.35^{+0.16}_{-0.13}$ & $-21.55^{+0.25}_{-0.23}$ & $5.18^{+3.27}_{-3.21}$ & $-0.12^{+0.33}_{-0.64}$  & $-4.39^{+3.70}_{-3.85}$ \\
    \hline
\end{tabular}    
\end{center}
\begin{flushleft}
    \begin{tabular}{ccc||ccc}
\hline 
    \multirow{3}{*}{$\bm{\langle m_\psi \rangle}$ ($10^{-22}$eV)} &  
       VADS Model&  $1.63\pm0.05$ & \multirow{3}{*}{$\bm{\langle r_a \rangle}$ (kpc)} & VADS Model&  ...\\
     & OM Model &  $1.30\pm0.18$ & & OM Model &  $1.71\pm0.32$\\
     & gOM Model &  $1.81\pm0.21$ & & gOM Model &  $5.30\pm1.12$\\
       \hline
    \hline
    \multirow{3}{*}{$\bm{\langle r_c \rangle}$ (kpc)}  &  
       VADS Model&  $0.50\pm0.01$ & \multirow{3}{*}{$\bm{\langle \beta_0\rangle}$}  & VADS Model&  ...\\
     & OM Model &  $0.59\pm0.05$ & & OM Model &  ...\\
     & gOM Model &  $0.46\pm0.04$ & & gOM Model & $0.12\pm0.08$\\
       \hline
       \hline
    & & & \multirow{3}{*}{$\bm{\langle \beta_\infty\rangle}$}  & VADS Model&  ...\\
     & &  & & OM Model &  ...\\
     & & & & gOM Model &  $-4.46\pm1.33$\\
       \hline
\end{tabular}
\end{flushleft}
    \caption{The median and the 68\% confidence interval of our posterior analysis for all the parameters. Regarding the boson mass $m_\psi$ (fourth row) we also derived averaged values (and its 68\% confidence interval) on all the galaxies and for each model of the anisotropy parameter taken into account. We report such averaged values below the main table.}
    \label{tab:posterior}
\end{table*}

We analysed the possibility of alleviating the tension on the value of the boson mass estimated using the kinematic of stars in dwarf galaxies and using cosmological data sets. Hence, we re-analysed the kinematics of stars in dwarf galaxies using a more complex model than the one used in \cite{Chen2017}, {\it i.e.} we took into account an anisotropic velocity distribution with the parameter $\beta(r)$ that varies with the distance from the centre of the galaxy.  Under this assumption, which founds its interest in the results obtained with ultra-faint dwarf galaxies in \cite{Hayashi2021}, we predicted the velocity dispersion along the line of sight using the equation \eqref{eq:sigmalos}, and we fit it to the kinematic data of eight dSph galaxies employing an MCMC algorithm. Since we adopted three different models for the $\beta(r)$ (more information is given in Sect. \ref{subsec:betas}), the number of free parameters varies depending on the specific model of $\beta(r)$. For the VADS Model, we have a two-dimensional parameter space $\bm{\theta} = [m_\psi, r_c]$. In the case of the OM Model, we have one more extra parameter $\bm{\theta} = [m_\psi, r_c, r_a]$. And, finally, for the gOM Model, we have five free parameters $\bm{\theta} = [m_\psi, r_c, r_a, \beta_0, \beta_\infty]$. The outcome of our analyses for each model and each dSph galaxy, {\it i.e.} the median value of each parameter and its 68\% confidence interval, is reported in Table \ref{tab:posterior}. 

\begin{figure}[!ht]
    \centering
    \includegraphics[width= \columnwidth]{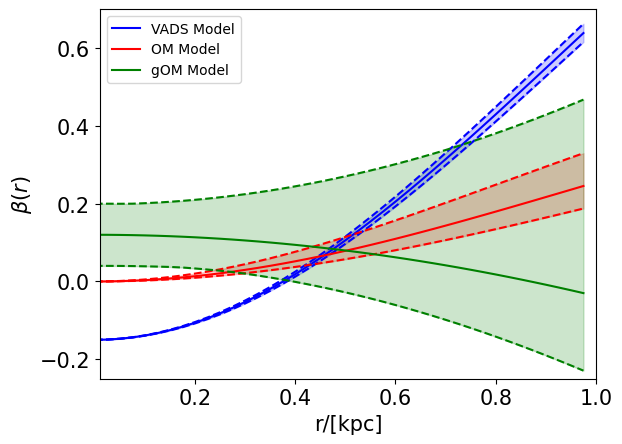}
    \caption{The figure depicts the radial profiles of the anisotropy parameter $\beta(r)$. In blue, red, and green are represented the three models used in our analysis ({\it i.e.}, the VADS, OM, and gOM models, respectively), and with solid lines their median profiles obtained with the median value of the parameters listed in Table \ref{tab:posterior}. Finally, shaded regions show the 68\% confidence interval around the median value.}
    \label{fig:betas}
\end{figure}

Although we ensured the convergence of each run of our MCMC analysis, the scale radius $r_a$ in OM and gOM models is poorly constrained. In fact, for both models, the whole range of prior values is covered within 99\% of the confidence interval. This is a fairly expected result, mainly because the observations only cover the innermost part of the galaxies up to 1 kpc. Nevertheless, if we look at the radial dependence of the anisotropy parameter constrained in the three models we found that in all models the velocity distribution tends to become more radially anisotropic except for the model gOM.  In Figure \ref{fig:betas}, we depict the radial behaviour of the anisotropy parameter and we show in blue, red, and green the VADS, OM, and gOM models, respectively. Solid lines represent the median profile of the anisotropy parameter obtained using the median values reported in Table \ref{tab:posterior}, and the dashed lines represent the $1-\sigma$ confidence intervals. The trend resulting from our analysis confirms, for the VADS and OM models, the one that has been recently highlighted in numerical simulations that solved the Schrödinger-Poisson equation for a boson with mass $0.8\times10^{-22}$ eV \cite{Dutta2023}. On the contrary, the gOM model, which was also used to fit the velocity dispersion of ultra-faint galaxies in \cite{Hayashi2021} and helped to relieve the tension with cosmological analysis, has a preference for a velocity distribution more tangentially anisotropic. This also contrasts results from numerical simulations of the dynamical heating in \cite{Dutta2023}. Nevertheless, it is worth mentioning that a velocity distribution more radially anisotropic is still allowed within 68\% of the confidence interval.

\begin{figure*}[!ht]
    \centering
    \includegraphics[width= 1.99\columnwidth]{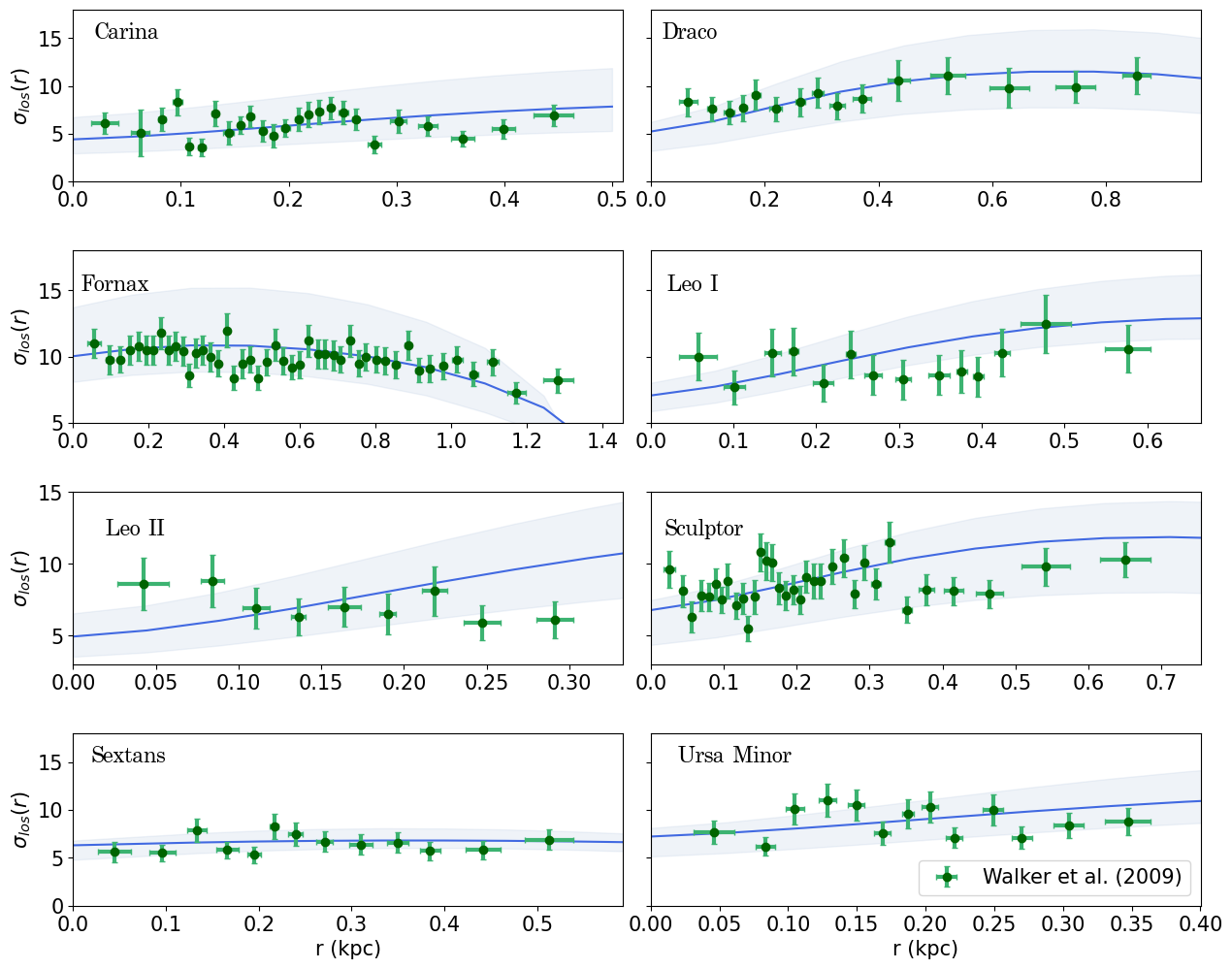}
    \caption{The figure depicts the radial velocity dispersion profile of the 8 dwarf galaxy Carina with a $\beta(r)$ given by the VADS model. In green are reported the data and their error bar, while the blue solid lines represent the median profiles obtained with the averaged median value of the parameters listed in Table \ref{tab:posterior}. Finally, shaded regions show the 68\% confidence interval around the median value.}
    \label{fig:profile_vads}
\end{figure*}

\begin{figure*}[!ht]
    \centering
    \includegraphics[width= 1.99\columnwidth]{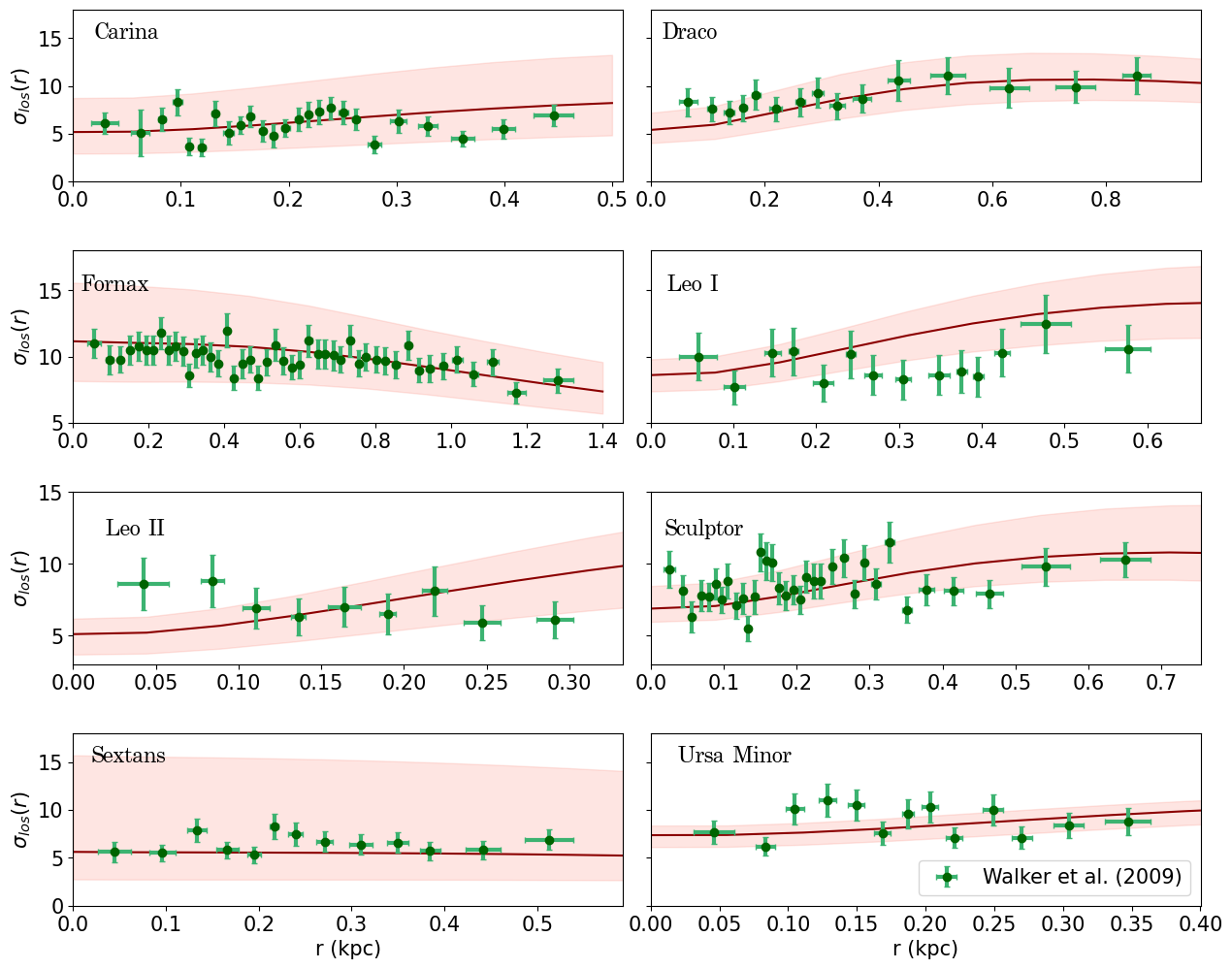}
    \caption{The same as Figure \ref{fig:profile_vads} but with a $\beta(r)$ profile given by the OM model. Here the averaged median model and the 68\% confidence interval are depicted in red and pink colors, respectively.}
    \label{fig:profile_OM}
\end{figure*}

\begin{figure*}[!ht]
    \centering
    \includegraphics[width= 1.99\columnwidth]{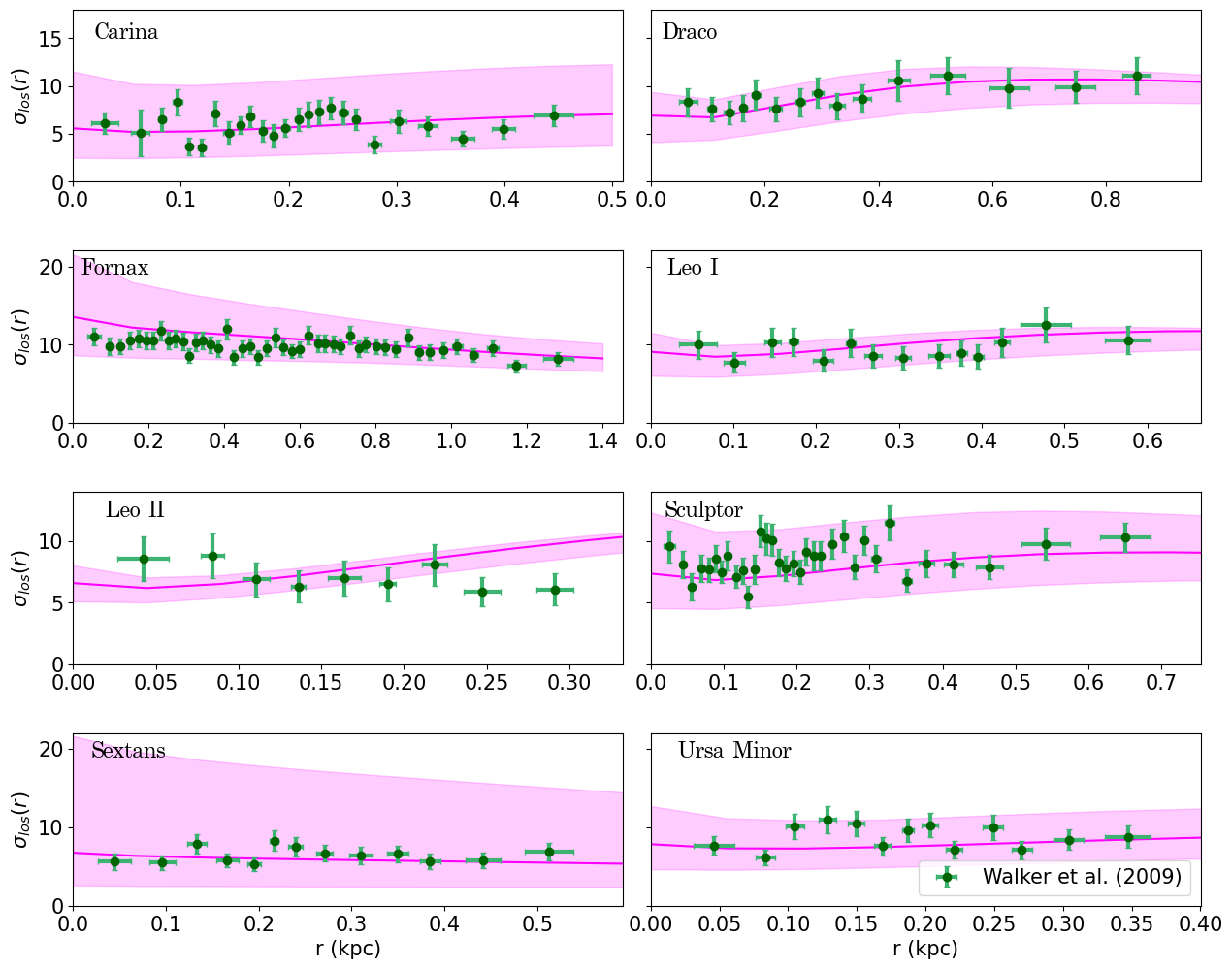}
    \caption{The same as Figure \ref{fig:profile_vads} but with a $\beta(r)$ profile given by the gOM model. In green, there are reported data and their error bar, while the magenta solid lines represent the median profiles obtained with the averaged median value of the parameters listed in Table \ref{tab:posterior}. Finally, shaded regions show the 68\% confidence interval around the median value.}
    \label{fig:profile_gOM}
\end{figure*}

On the other hand, looking at each galaxy, both the median values of boson mass $m_\psi$ and of the solitonic core radius $r_c$ of all the anisotropic velocity distribution models agree with each other within $1-\sigma$. Interestingly, for each model, the averaged values over different galaxies of the boson mass and the solitonic core radius cluster around two values: ${\langle m_\psi \rangle} (10^{-22} = [1.63\pm0.05; 1.30\pm0.18; 1.81\pm0.21]$ eV and $\langle r_c \rangle (\text{kpc}) = [0.50\pm0.01; 0.59\pm0.05; 0.46\pm0.04]$ for  VADS, OM, and gOM models, respectively. In Figures \ref{fig:profile_vads}, \ref{fig:profile_OM}, and \ref{fig:profile_gOM}, we show the effectiveness of such a median DM halo to fit the velocity dispersion of the eight dSph galaxies adopting the VADS, OM, and gOM model, respectively. In all Figures, the green points represent the observed velocity dispersion data with their statistical uncertainties, while the solid lines and the shaded regions depict the velocity dispersion predicted using the averaged values reported in Table \ref{tab:posterior}. The trend of the velocity dispersion is always very well reproduced, except for the galaxy Leo II where the average model predicts a lower velocity dispersion in the first few hundred pc of the galaxy. However, it is rather expected that the averaged model whose solitonic radius is roughly two times the half-light radius of  Leo II, cannot correctly predict its inner velocity dispersion. Nevertheless, it is important to note that, within the 95\% confidence interval, the averaged model would agree with the data of the velocity dispersion even in the innermost part of the galaxy.  Finally, such results confirm the previous findings in \cite{Chen2017} that one single boson mass of $\sim 10^{-22}$  could
effectively explain the kinematics of stars in dSph galaxies. However, unlike what happens in ultra-diffuse galaxies \cite{Hayashi2021},  in our analysis, the preferred value is still  $\sim 10^{-22}$ eV. Therefore,  assuming an anisotropy parameter that varies with distance from the centre of the galaxy neither eliminates nor reduces the tension between the preferred value of the boson mass at the galactic and at cosmological scales.


This result points to different scenarios:  (1) the simplifying assumption of spherical symmetry adopted in the kinematic model of the stars is not adequate since real dwarfs appear elliptical on the sky \cite{Irwin1995,Salomon2015}; (2)  the wave DM model made of a single not self-interacting boson cannot correctly describe the observations at all scales. Hence, this scenario would be ruled out, but this would not be the case for the wave DM model itself. Indeed, there is still the possibility of the creation, in the primordial Universe, of more than one ultralight particle with different masses and, hence, capable of dominating the evolution of matter on different astrophysical and cosmological scales. 

The first scenarios should be explored. However, data does not allow us to investigate how the simplifying assumptions of spherical symmetry, that we have adopted, can affect our final results. In the future, we plan to explore this scenario using synthetic proper motion data mimicking next-generation astrometric missions such as \textit{Theia} \cite{Malbet2021}.

The second scenario is more intriguing. In the String Axiverse \cite{Svrcek2006,Arvanitaki2010,Cicoli2012,Cicoli2022},  the existence of multiple axions or axion-like particles with a wide mass spectrum is naturally predicted. The different particles may simultaneously act like wave DM candidates. In fact, there are already some studies considering a multi-component scenario and its impact on the formation and evolution of the galaxies. For instance, in \cite{Luu2020}, a Jeans analysis of four ultra-faint galaxies, namely Segue I, Carina II, Reticulum II, and Hydrus I, suggested the possibility of the existence of an additional axion with mass $\simeq 10^{-20}$ eV, and of a third axion with mass $\gtrsim 10^{-18}$ eV. Such a possibility would also agree with the Jeans analysis of \cite{Hayashi2021} that indicates Segue I is well described by a boson with mass $\simeq 10^{-19}$ eV. Additionally, in \cite{Tellez2022}, cosmological data such as Baryon Acoustic Oscillation, Supernovae, Lyman-$\alpha$ forest, and Big Bang Nucleosynthesis have been used to probe the multi-component scenario. The results therein suggest a second boson of mass in agreement with the aforementioned galactic studies that would not alter the cosmological dynamics which would be still dominated by the lighter boson.

Finally, our analysis excludes the possibility of alleviating or removing the existing tension on the favoured value of the boson mass between constraints coming from the kinematics of galaxies and those coming from cosmological data. Although, following the analysis with ultra-faint galaxies in \cite{Hayashi2021}, we have considered an anisotropy parameter varying with the distance from the centre of the galaxy, the favoured value of the boson mass is still of the order of $\sim 10^{-22}$ eV. Contrary, in ultra-faint galaxies, this assumption favours a heavier boson with a mass of the order of $\simeq 10^{-19}$ eV. Therefore, if we want to retain the wave DM paradigm, we can argue that a multi-flavoured axion component is needed to resolve the tension on the boson mass.

\section{Conclusions}\label{sec:concl}

The Wave DM scenario has made a comeback in the last decades as an alternative to the cold dark matter model. The need arises from the problems that the latter encounters on a galactic scale such as the cusp-core problem, the missing satellite problem, and the too-big-to-fail problem, among others. The wave DM model can easily and naturally explain some of these problems. The latter foresees, on the one hand, the existence of a solitonic core in the innermost part of each virialized halo and, on the other hand, a cut-off in the matter power spectrum that would prevent the formation of DM halos that are too small to form galaxies. However, the wave DM model is unable to explain both the kinematics of galaxies and the formation of structures on a cosmological scale with a single particle. While the former prefers a boson with a mass of the order of $\sim 10^{-22}$ eV, the latter prefers one with a mass of the order of $\sim 10^{-20}$ eV. { Despite the aforementioned difficulties,  the comoving cumulative stellar mass density of massive galaxies at redshift $z=8$ and $z=9$ measured by the JWST may be explained within the wave DM framework with a boson of $\sim 10^{-22}$ eV. However, this solution encounters difficulties in explaining the same data at redshift $z=10$ \cite{Gong2023}.}

In ultrafaint galaxies, this tension can be alleviated by modelling the kinematics of stars including a velocity distribution with an anisotropy parameter that varies with distance from the centre of the galaxy. In our analysis, we investigated the possibility that this assumption could alleviate or eliminate the tension existing in the favoured value of the boson mass also in the case of dSph galaxies. However, our results show that this is not the case and that, even if we introduce this complication into the kinematic model, the favoured value of the boson mass remains of the order of $\sim 10^{-22}$ eV. One may argue that the single flavour axion scenario would be ruled out, but the multi-component scenario that naturally arises in the String Axiverse, where one may have a heavier boson explaining the dynamics of dwarf galaxies and another boson explaining the cosmological evolution as recently suggested in \cite{Pozo2023}, is still possible and must be carefully explored. { For instance, on one hand, its predictive power can be affected by the existence of degeneracies between the masses of the two species. On the other hand, several mass ratios are found to fit data as well as the standard $\Lambda$CDM model with no significant changes on the CMB power spectrum for scalar field masses higher than $\sim 10^{-26}$ eV \cite{Tellez2022},  but serious departures with respect the matter power spectrum of CDM can appear if the less massive particle would dominate the cosmological evolution.}

\vspace{1cm}

\textit{Acknowledgements}  ---  IDM acknowledges support from the grant PID2021-122938NB-I00 funded by MCIN/AEI/10.13039/501100011033 and by “ERDF A way of making Europe”. 

\bibliography{biblio}
\end{document}